\documentstyle{elsart}

\begin{document}
{\small \hfill  INP~MSU-99-23/581, TTP99-27, 16 June 1999.}
 
\begin{frontmatter}
\title{Advanced method of solving recurrence relations
for multi-loop Feynman integrals. \thanksref{paper}}

\thanks[paper]{Talk presented at the AIHENP'99 (Heraklion, Greece, April
12-16, 1999)}

\author[A]{P.\,A.\,BAIKOV \thanksref{AA}}

\thanks[AA]{Supported in part by INTAS (grant YS-98-174),
Volkswagen Foundation (grant No.~I/73611)
and RFBR (grant 98-02-16981);
e-mail:baikov@theory.npi.msu.su}

\address[A]{Institute of Nuclear Physics, Moscow State University,
Moscow~119899, Russia}

\begin{abstract}
The systematic approach to solving the
recurrence relations for multi-loop integrals is described.
In particular, the criteria of their reducibility is
suggested.
\end{abstract}

\end{frontmatter}

\section{Introduction.}

The integration by part method~\cite{ch-tk} is a very convinient
and in most case unique opportunity to evaluate higher order
perturbative corrections in 
quantum field theory. It provides with the relations connecting Feynman
integrals with different powers of their denominators. In many
cases it is possible to construct the recursive procedure which
expresses an integral with given degrees of the denominators as
a linear combination of a few so-called master integrals.

The construction of such procedure is a nontrivial problem
even at two-loop level~\cite{Tar97}. At three loops the case of
vacuum integrals with one non-zero mass and various numbers of
massless lines has been considered~\cite{REC}.  At four loops 
the recursive procedure was constructed only for
some particular cases~\cite{4loop}.

The recursive procedures mentioned above were constructed as
the result of the "hand-work" combining of original
relations.  The total amount of this "hand-work" grows in very large
extent with number of loops and starting from 4-loop level
became unmanageable. Although there is some practical
experience on that (unfortunately, not properly described in the
literature), the whole approach is hard to be made automatic.

From the other side, recently the systematic approach to solving
the recurrence relations for multi-loop integrals was suggested
\cite{ES}, see also \cite{BS}. In this paper we describe the further
progress in this direction.

\section{Integral representation}

The basic ideas of the approach is the following. Suppose we
need to solve the relations for the objects
$B(\underline{n})\equiv B(n_1,...,n_N)$:

\begin{equation}
R(I^-,I^+) B(\underline{n})=0, \label{recrel}
\end{equation}
$$I^-B(n_k)=B(n_k-\delta_{ki}), \quad I^+ B(n_k)=n_i B(n_k+\delta_{ki}).$$

\noindent
that is represent the arbitrary $B(\underline{n})$ as the sum of the
irreducible objects $B(\underline{n}_k)$, where
$\underline{n}_k, k=1,...,M$ are specific sets of index values
(usually "zeros" and "ones"):

\begin{equation}
B(\underline{n})=\sum_k c^{k}(\underline{n})
B(\underline{n}_k),
\label{gensol}
\end{equation}

\noindent
Note that coefficient functions $c^{k}(\underline{n})$ are
linear independent solutions of (\ref{recrel}).(Proof: act by
(\ref{recrel}) on (\ref{gensol}). If $B(\underline{n}_k)$
irreducible, than $R(I^-,I^+) c^{k}(\underline{n})=0$ for all
$k$.)

\noindent
By construction, the coefficient functions fulfil the initial
conditions

\begin{equation}
c^{i}(\underline{n}_k)=\delta_{ik}.
\label{icond}
\end{equation}

If we have some set of solutions $f^{(k)}(\underline{n})$, 
we can construct a desirable set $c^{k}(\underline{n})$ as their linear
combinations, which fulfil (\ref{icond}). Let us try the following
complex-integral representation:

\begin{equation}
f^{k}(\underline{n})=\int \frac{dx_1...dx_N}{x_1^{n_1}...
x_N^{n_N}}g(\underline{x}).
\label{sol}
\end{equation}

Acting by (\ref{recrel}) on (\ref{sol}) and using integration by
part over $x_i$ we got:

\begin{equation}
R(I^-,I^+) f^{k}(\underline{n})=\int
\frac{dx_1...dx_N}{x_1^{n_1}... x_N^{n_N}} R(x_i,\partial /
\partial x_i) g(\underline{x}) + (\mbox{surface terms}).
\nonumber
\end{equation}

\noindent
(Due to the integration by part the order of operators in $R$
changes to reverese one). Thus, choosing $R(x_i,\partial / \partial
x_i) g(\underline{x}) =0$ and removing surface terms by proper
choice of integration contours we can fulfil the (\ref{recrel}).

Suppose we are interesting in the solution which should be equal
to zero if $n_i<1$ for some $i$ (for example, if the Feynman
integral vanishes when degree of line number $i$ is
non-positive). In this case we can choose the integration
contour for $x_i$ as small circle around $x_i=0$ and the integration will
lead to the calculation of $(n_i-1)^{\mbox{th}}$ Taylor coefficient in
$x_i$.
In the following we will call such $n_i$ as "Taylor" type recurrence
parameters.

Unfortunately, such "pure" case is not very practical. Instead of that,
we ussually meet with "mixed" Taylor case, when Feynman integral
vanishes if one shrinks some set of lines. Lets us explain, how this
"mixed" case can be decomposed into the sum of "pure" cases.

\section{Combinatorial decomposition}

Let us consider first the simple example. Suppose we need to
calculate $f_{n_1 n_2}$ (two-parameter recurrence problem), with
additional condition $f_{n_1 n_2}=0$ if $n_1\leq0$ and
$n_2\leq0$. Let us define projectors $I_i, O_i$:

$$I_i f_{n_1 n_2}=(f_{n_1 n_2}\ \mbox{if}\ n_i>0\ \mbox{else}\
0),\ O_i=1-I_i,\ i=1,2.$$

In these notations the condition on $f_{n_1 n_2}$ reads

\begin{equation}
O_1 O_2 f_{n_1 n_2}=0. \label{cond}
\end{equation}

Then $f_{n_1 n_2}$ can be decomposed in the following way (in
the second equality we omit $O_1 O_2$ contribution):

\begin{eqnarray*}
f_{n_1 n_2}&=&(I_1+O_1)(I_2+O_2)f_{n_1 n_2}= (I_1 I_2 + I_1
O_2 + O_1 I_2)f_{n_1 n_2}=\nonumber \\
&=& (I_1 + I_2 - I_1 I_2)f_{n_1 n_2}.\nonumber
\end{eqnarray*}

That is the original recurrence problem is represented as
algebraic sum of the following sub-problems: the first when we
neglecting all $f_{n_1 n_2}$ with $n_1$ non-positive, the second
neglecting with $n_2$ non-positive, and the third when
$n_1$ or $n_2$ is non-positive.  Each of these sub-problems has
"Taylor" type recurrence parameters, which significantly
simplify the representation (\ref{sol}).

In the general case, we will have the list of conditions of
(\ref{cond}) type; each item of this list corresponds to some
case when the Feynman integral vanishes if one shrinks the given
set of lines.  To get the decomposition in general case, we
need to calculate $\Pi_i (I_i+O_i)$, omit all monoms
which include as sub-monom the item from the list, and in the rest ones
make
the substitution $O_i=1-I_i$.  (Note, that one can obtain
various versions of this decomposition by adding "zero" monoms
from the list with arbitrary coefficients, but we found more
practical the recipe described above).

As next example let us consider the 2-loop
massless propagator type diagrams.

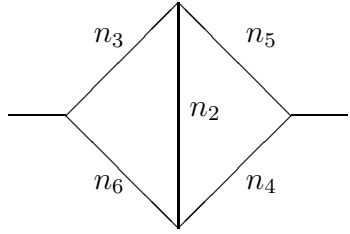
\setlength{\unitlength}{0.15mm}
\begin{center}
\begin{picture}(300,200)
\put(50,100){\line( 1, 1){100}}
\put(50,100){\line( 1,-1){100}}
\put(250,100){\line(-1,-1){100}}
\put(250,100){\line(-1, 1){100}}
\put(0,100){\line( 1, 0){50}}
\put(250,100){\line( 1, 0){50}}
\put(150,200){\line( 0, -1){200}}

\put(75,165){$n_3$}
\put(75,35){$n_6$}
\put(210,35){$n_4$}
\put(210,165){$n_5$}
\put(20,105){\phantom{$n_1$}}
\put(160,100){$n_2$}
\end{picture}
\end{center}

The "zero" list consists of $\{O_3 O_5, O_3 O_2, O_2 O_5, O_4
O_6, O_2 O_6, O_4 O_2, O_3 O_6, O_4 O_5\}$. The procedure
described above leads to decomposition

\begin{equation}
f_{n_2...n_5}= (I_3 I_4 I_5 I_6 + I_2 I_3 I_4 + I_2 I_5 I_6 - 2
I_2 I_3 I_4 I_5 I_6) f_{n_2...n_5}.
\label{dec5}
\end{equation}

The representation (\ref{sol}) in this example will read

\begin{eqnarray}
f_{n_2...n_5}&=&
\int\frac{dx_2...dx_6}{x_2^{n_2}...x_6^{n_6}}
{P(x_1,\dots,x_6)^{D/2-2}}
\,,
\label{sol5}
\end{eqnarray}
\begin{eqnarray*}
P&=&
(x_1+x_2)(x_1x_2-x_3x_4-x_5x_6)+(x_3+x_4)(x_3x_4-x_1x_2-x_5x_6)
\nonumber\\&+&
(x_5+x_6)(x_5x_6-x_1x_2-x_3x_4) +x_1x_3x_6+x_1x_4x_5+x_2x_3x_5
+x_2x_4x_6,
\end{eqnarray*}

\noindent 
where $x_1$ corresponds to external line. Since the dependence
on the external momentum is trivial, in the following we can set
$x_1=1$ without loss of generality (the strict proof can be done
by auxiliary integration over external momentum $q^2$ with
weight $1/(q^2-1)$).

Let us now apply to (\ref{sol5}) the decomposition (\ref{dec5}).
The first term in (\ref{dec5}) implies the Couches integration
over $x_3, x_4, x_5, x_6$, which leads to Taylor expansion in
these variables. As the result we get a set of
integrals of the following type:

\begin{eqnarray}
\int
\frac{dx_2}{x_2^{n_2}}
[x_2(x_2+1)]^{D/2-2-c}
\propto
(-1)^{n_2-c}\frac{(D/2-1)_{-c}(D/2-1)_{-c-n_2}}{(D-2)_{-2c-n_2}}\nonumber
\end{eqnarray}

The second term in (\ref{dec5}) leads to Taylor expansion in
$x_2,x_3,x_4$, the remaining integrals are:

\begin{eqnarray}
\int
\frac{dx_5dx_6}{x_5^{n_5}x_6^{n_6}}
[x_5x_6(1-x_5-x_6)]^{D/2-2-c}
\propto\frac{(D/2-1)_{-n_5-c}(D/2-1)_{-n_6-c}}{(3D/2-3)_{-n_5-n_6-3c}}
\nonumber
\end{eqnarray}

\noindent The third term leads to similar contribution (with
substitutions $3\leftrightarrow 5,4\leftrightarrow 6$).
Finally, the forth term leads to zero contribution (the polinom
$P$ vanish if $x_2=x_3=x_4=x_5=x_6=0$).

Let us consider this last sub-case more attentively. According
to previous definitions, in the corresponding recurrence
sub-problem we should omit any integral with $n_i\leq0$ for some
$i$, that is at least with one line (with label $i$) shrinked.
The zero answer means that recurrence procedure re-express the
integrals with all $n_i>0$ (in particular with all $n_i=1$)
through the integrals with at least one line shrinked.

As the result, we get the necessary condition of reducibility:
the given integral with $n_i>0, i\in S$ can be expressed through
more simple integrals (with some line shrinked) only if the Couches
integration over the corresponding $x_i, i\in S$ in (\ref{sol})
leads to zero result:

\begin{equation}
\Pi_{i\in S} I_i f(\underline{n})=0.
\label{mcond}
\end{equation}

\noindent
Note, that in more complicated cases it may happen that the
integrand does not vanish after Taylor expansion and the zero
result can appear after remaining integration; we mean this
final zero.

Moreover, it looks like (\ref{mcond}) is at the same
time the sufficient condition. The formal proof requires consideration of
some pathological cases and still not completed.
From the other hand, in practice we should construct
the reduction procedure explicitly.
In cases we met up to now (up to 3-loop propagator type massless
integrals), if (\ref{mcond}) takes place, it appeared to be possible.
We will describe it in the expanded version of this paper.

\section{Conclusion.}
In this paper we shortly describe the systematic approach to solving the
recurrence relations for multi-loop integrals.
In particular, the criteria of their reducibility is suggested.
We believe that this approach will allow to make automatic the solving
procedure which help in the practical calculation of such integrals.

The author would like to thank J.H.K\"uhn, K.Chetyrkin and K.Melnikov
for useful discussions.

\end{document}